\documentclass[prl,twocolumn,showpacs,superscriptaddress]{revtex4}

\usepackage{graphicx}
\usepackage{mathrsfs}
\usepackage{dcolumn}
\usepackage{bm}
\usepackage{amsmath}
\usepackage{amsfonts}
\usepackage{color}
\usepackage{epstopdf}

\newcommand{\ket}[1]{\left|#1\right\rangle}

\newcommand{\up}{\uparrow}
\newcommand{\dn}{\downarrow}
\newcommand{\bk}{\mathbf k}

\newcommand{\beginsupplement}{%
        \setcounter{table}{0}
        \renewcommand{\thetable}{S\arabic{table}}%
        \setcounter{figure}{0}
        \renewcommand{\thefigure}{S\arabic{figure}}%
     }

\begin{document}

\title{Magnetic semimetals and quantized anomalous Hall effect in EuB$_6$}

\author{Simin Nie}
\affiliation{Beijing National Laboratory for Condensed Matter Physics,
and Institute of Physics, Chinese Academy of Sciences, Beijing 100190, China}
\affiliation{Department of Materials Science and Engineering, Stanford University, Stanford, California 94305, USA}
\author{Yan Sun}
\affiliation{Max Planck Institute for Chemical Physics of Solids, N\"othnitzer Str. 40, 01187 Dresden, Germany}
\author{Fritz B. Prinz}
\affiliation{Department of Materials Science and Engineering, Stanford University, Stanford, California 94305, USA}

\author{Zhijun Wang}
\email{wzj@iphy.ac.cn}
\affiliation{Beijing National Laboratory for Condensed Matter Physics,
and Institute of Physics, Chinese Academy of Sciences, Beijing 100190, China}
\affiliation{University of Chinese Academy of Sciences, Beijing 100049, China}

\author{Hongming Weng}
\affiliation{Beijing National Laboratory for Condensed Matter Physics,
and Institute of Physics, Chinese Academy of Sciences, Beijing 100190, China}
\affiliation{University of Chinese Academy of Sciences, Beijing 100049, China}

\author{Zhong Fang}
\affiliation{Beijing National Laboratory for Condensed Matter Physics,
and Institute of Physics, Chinese Academy of Sciences, Beijing 100190, China}
\affiliation{University of Chinese Academy of Sciences, Beijing 100049, China}

\author{Xi Dai}
\email{daix@ust.hk}
%\affiliation{Beijing National Laboratory for Condensed Matter Physics,
%and Institute of Physics, Chinese Academy of Sciences, Beijing 100190, China}
\affiliation{Department of Physics, Hong Kong University of Science and technology, Clear Water Bay, Kowloon, Hong Kong}

\date{\today}

\begin{abstract}
Exploration of the novel relationship between magnetic order and topological semimetals has received enormous interest in a wide range of both fundamental and applied research. Here we predict that ``soft" ferromagnetic (FM) material EuB$_6$ can achieve multiple topological semimetal phases by simply tuning the direction of the magnetic moment.
Explicitly, EuB$_6$ is a topological nodal-line semimetal when the moment is aligned along the [001] direction, and it evolves into a Weyl semimetal with three pairs of Weyl nodes %(one pair around each Z point)
by rotating the moment to the [111] direction. %, resulting in large Chern number in the (111) plane in momentum space.
Interestingly, we identify a novel semimetal phase featuring the coexistence of a nodal line and Weyl nodes with the moment in the [110] direction. Topological surface states and anomalous Hall conductivity, which is sensitive to the magnetic order, have been computed and are expected to be experimentally observable. %Due to three pairs of Weyl nodes of the [111] magnetism,
Large-Chern-number quantum anomalous Hall effect can be realized in its [111]-oriented quantum-well structure.
\end{abstract}

\maketitle

\begin{figure}[t]
 %\captionstyle{centerlast}
%\includegraphics[clip,scale=0.17, angle=0]{wzj1}
\includegraphics[clip,scale=0.30, angle=0]{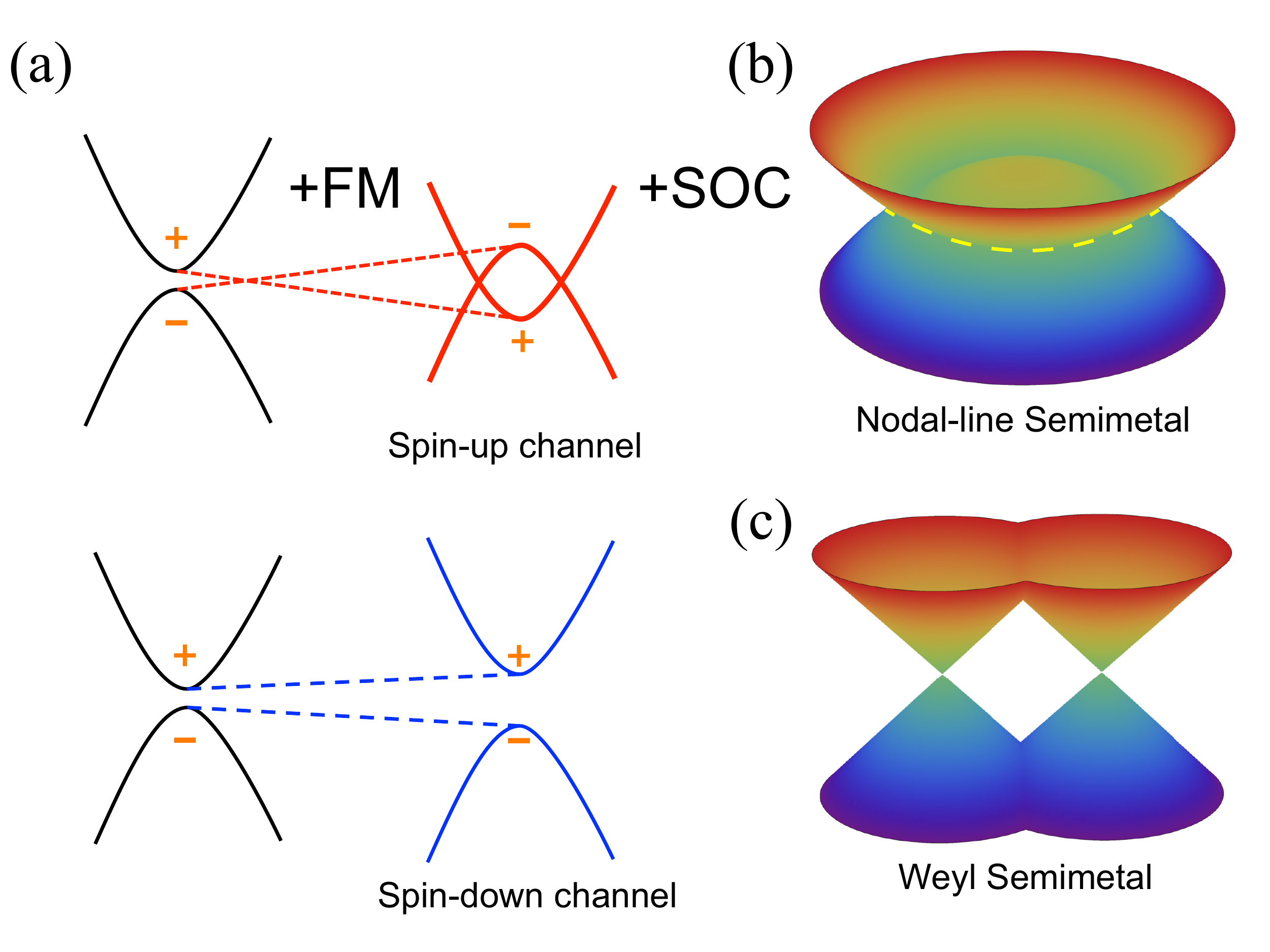}
\caption{~$\boldsymbol{|}$ \textbf{Schematics of the band inversion with opposite exchange splitting in FM order.}
Starting from the narrow-gap centro-symmetric semiconductor with even-parity conduction band and odd-parity valence band, by introducing FM, the exchange splitting pushes the spin-up valence band upwards, while it pushes the spin-up conduction band downwards. %because they exhibit opposite effective exchange splitting.
This results in band inversion in the spin-up channel (the upper panel of \textbf{(a)}) ,
but it enlarges the band gap in the spin-down channel (the lower panel of \textbf{(a)}).
After considering SOC, the system becomes either a TNLS \textbf{(b)} %with mirror symmetry,
or a magnetic WSM \textbf{(c)} with anomalous Hall effect.}
 \label{fig:1}
\end{figure}

Topological semimetals exhibit topologically protected quasiparticles near the Fermi
level, among which Dirac fermions \cite{young2012dirac,wang2012dirac,wang2013three}
and Weyl fermions \cite{nielsen1983adler,wan2011topological,balents2011weyl,xu2011chern} have
elementary particle counterparts in high-energy physics while others (such as threefold spin-1 fermions \cite{bradlyn2016beyond,tang2017,zhang2018}, nodal-line
fermions \cite{burkov2011topological,fang2015topological,weng2016topological,nie2018topological,Yu2019}) not.
These quasiparticles are classified according to the degeneracy and the shape of the
band-touching points. The discrete band-touching points with fourfold (twofold)
degeneracy are termed Dirac (Weyl) fermions, while the line-contact band-touching
points are known as nodal-line fermions. By breaking certain crystalline symmetry (such as mirror
symmetry), topological nodal-line semimetals (TNLSs) can be driven into Weyl semimetals (WSMs) \cite{yuHfC}, Dirac semimetals (DSMs)  \cite{kim2015dirac,yu2015topological}, topological insulators (TIs) \cite{srIo3,Nie03102017,tase3nie}.
As we know, for centro-symmetric systems with time-reversal symmetry (TRS), the band inversion between two bands with opposite parity,
happening at a single time reversal invariant momentum (TRIM), results in a TNLS/TI  in the absence/presence of spin-orbit coupling (SOC) \cite{xu2017,huang2016}.
However, for those with magnetic order (breaking TRS),
the band inversion between the two single-degenerate bands with opposite parity can give rise to $\chi=1$ with the definition:
\begin{equation}
(-1)^\chi\equiv\prod_{j=\{1,2,\cdots,n_{occ}\},~\Gamma_i=\text{TRIMs}} \xi^j_{i}
\end{equation}
where $\xi_i^j$ is the parity eigenvalue of the $j$-th band at the TRIM $\Gamma_i$, and $n_{occ}$ is the total number of the occupied bands.
We note that $\chi=1$ implies that the system \emph{cannot} be fully gapped \cite{Hughes2011, Wang2016} even with SOC.
Generally speaking, it can be either a TNLS or a WSM, as shown in Fig. 1, depending on the symmetry with magnetism.

Therefore, magnetic order provides us %an alternative and
a promising way to control the symmetry and topology. For example, new types of DSMs are proposed to exist in the antiferromagnetic materials \cite{tang2016dirac,HuaDirac}. Compared with ``hard" magnetic materials, ``soft" magnetic materials facilitate the tuning of magnetic order, providing an ideal platform for
exploring the interplay between magnetic orders and topological semimetals.

Over the decades, europium hexaboride (EuB$_6$), a well-known ``soft" magnetic
material, has been extensively studied due to the appearance
of interesting electrical transport properties near the FM
transition temperatures, such as the metal-insulator transition \cite{GUY19801055,
nyhus1997}, the giant blue shift of the unscreened plasma frequency
\cite{degiorgi1997low,broderick2002scaling}, the large zero-bias anomalies
\cite{amsler1998electron}, large negative magnetoresistance \cite{GUY19801055,
sullow2000metallization}, etc. At T$_{c1}=15.3$ K \cite{sullow1998structure,
brooks2004magnetic}, a phase transition from the paramagnetic (PM) phase to the
FM phase with moment oriented to [001] direction (called FM1) is experimentally
observed in EuB$_6$,  along with a drop of an order of magnitude in its resistivity \cite{GUY19801055}.
Evidence of another phase transition from FM1 to a new FM phase with the moment
oriented to the [111] direction (called FM2) is observed at T$_{c2}=12.5$ K \cite{sullow1998structure, brooks2004magnetic}.
Recent Andreev reflection spectroscopy reported that only about half of the carriers are spin polarized at the Fermi level (E$_\text{F}$)~\cite{zhang2008spin}, which seems to be in contradiction with the previous first-principles calculations suggesting a half-metallic ground state at the charge neutrality level~\cite{massidda1996electronic, kunevs2004kondo}. The incompatible results could be explained by the change of chemical potential
due to the deficiencies of samples. %So far,
Although there are several
theoretical calculations, the exploration of the topological properties
has not been reported, which might shed light on the explanation of the above
electrical transport properties. % such as the large zero-bias anomalies \cite{amsler1998electron} and large negative magnetoresistance \cite{GUY19801055,sullow2000metallization}.

%{\color{red}
In this work, %based on first-principle calculations,
we have systematically investigated the electronic structures of EuB$_6$ in both PM and FM phases. %tuned by temperature or an external magnetic field.
%By treating the localized $f$ states of Eu as core states to simulate the PM state,
%a combination of modified Becke-Johnson (mBJ) exchange potential with local density
%approximation calculation
We show that PM EuB$_6$ is an intrinsic semiconductor [Fig.~2(c)] with a tiny gap (about 20 meV) at three Z points (including X, Y and Z points), which is in good agreement with the experimental observations of the semiconductor behavior at high temperature \cite{eub6arpes}.
Once the temperature is below FM T$_c$,
%it goes through two consecutive FM transitions.
%Our calculations show that, due to
the consequent magnetic moment has opposite effective exchange splitting on the low-energy bands, which leads to the band inversion at three Z points in the spin-up channel, but enlarges the band gap in the spin-down channel [as shown in Fig.~1(a) or Fig.~2(d)].
%The resulting half-metallic band structure is consistent with the observed drop of an order of magnitude in its resistivity \cite{GUY19801055}.
%The band inversion indicates nontrivial band topology in FM EuB$_6$, which is still unrevealed in the literature.
The band inversion between two bands with opposite parity resulting in $\chi=1$ represents an ideal toy model, guaranteeing the existence of nodal lines or odd pairs of Weyl points in centro-symmetric materials \cite{Hughes2011, Wang2016}.
As expected, in the FM1 phase, %below T$_{c1}$,
EuB$_6$ is a TNLS with three nodal lines (one for each Z point) protected by mirror symmetry $\hat{M}_z$, while %below T$_{c2}$,
it is driven into a WSM with three pairs of Weyl nodes (one pair for each Z point) in the FM2 phase. %near E$_\text{F}$. %at the Fermi level (E$_F$).
Interestingly, by rotating the magnetic moment to the [110] direction (called FM3), a novel phase with coexistence of a nodal line and Weyl nodes is found. Topological surface states and anomalous Hall conductivity (AHC) are obtained.
% (\ie drumhead states for nodal lines or arc states for Weyl nodes) are obtained. %by our calculations.
The computed AHC suggests that FM2 and FM3 phases exhibit substantial anomalous Hall effect due to the existence of Weyl nodes, while the AHC of the FM1 phase is almost zero.
%Because three pairs of Weyl nodes in FM2 EuB$_6$ are related by the threefold rotation symmetry about the [111] direction,
In addition, large-Chern-number quantum anomalous Hall effect (QAHE) is proposed to be realized in its [111]-oriented quantum-well structure.

\begin{figure}[t]
 %\captionstyle{centerlast}
%\includegraphics[clip,scale=0.17, angle=0]{wzj2}
\includegraphics[clip,scale=0.18, angle=0]{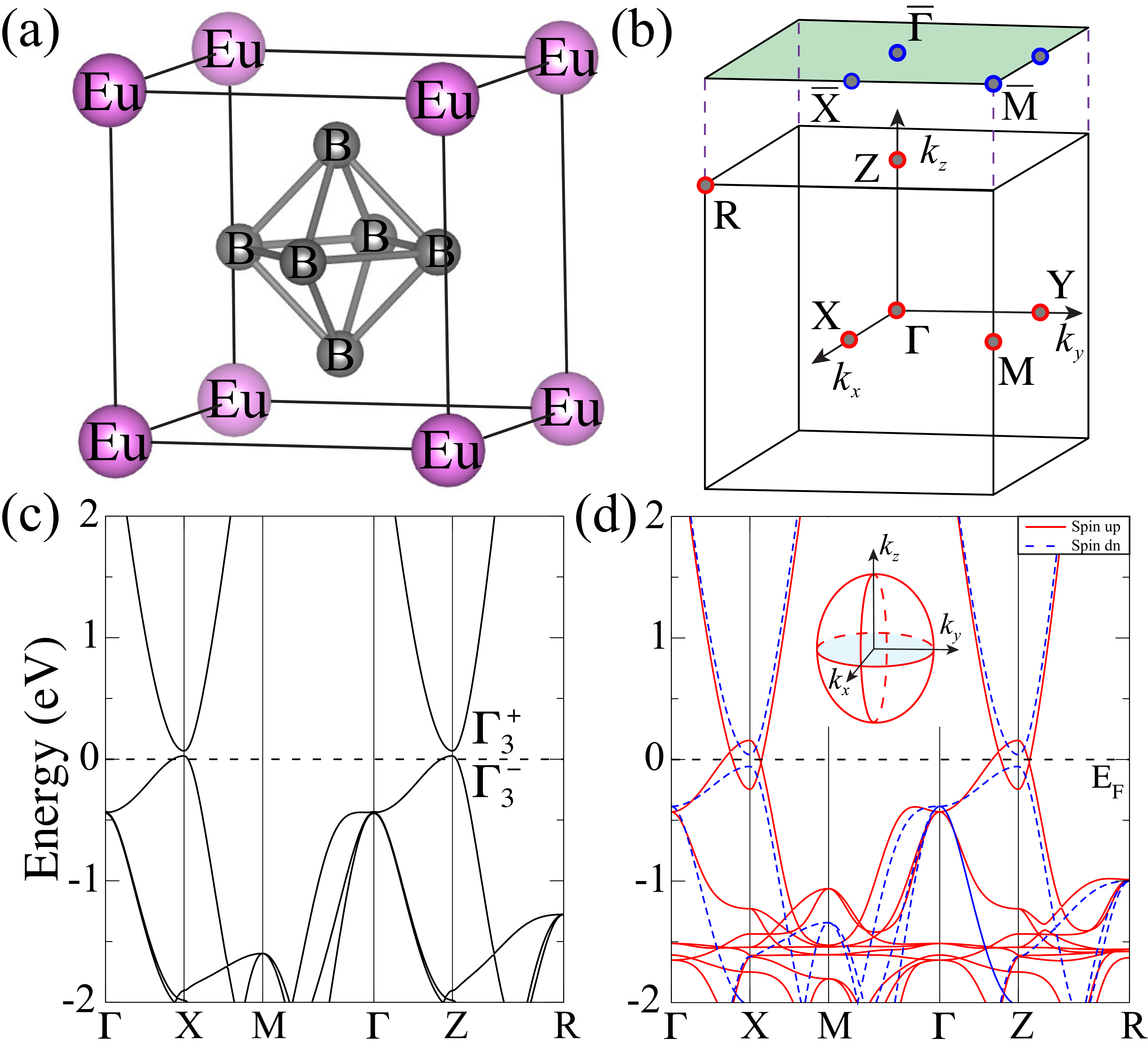}
\caption{~$\boldsymbol{|}$ \textbf{Crystal structure and band structures of EuB$_6$.} \textbf{(a)} Crystal structure of EuB$_6$. \textbf{(b)} The bulk and surface Brillouin zones (BZs)
for EuB$_6$. \textbf{(c)} mBJ band structure of EuB$_6$ with $``\text{open core}"$
treatment. \textbf{(d)} GGA+U band structure of EuB$_6$. The spin-up and spin-down bands
are colored in red and blue, respectively. %It shows that it's metallic in the spin-up channel, but gapped in the spin-down channel.
The inset schematically shows three of the five intersecting nodal lines at Z point.
} \label{fig:2}
\end{figure}

\emph{Crystal structure and symmetries.}
EuB$_6$ crystallizes in the CsCl-type structure \cite{TARASCON1981133} of space group Pm3m (No. 221), as shown in Fig. 2(a).
Eu and B atoms occupy the $1a$ (0, 0, 0) and $6f$ (1/2, 1/2, $u$)
Wyckoff positions, respectively. In the center of the cube with Eu atoms at the corners, six B atoms
form an octahedral cage. %, where the intra- and inter-octahedron B-B distances equal to $d_{\text{intra}}=\sqrt{2}a (1/2-u)$ and $d_{\text{inter}}=2 a u$, respectively.
The optimized lattice constant $a$ and internal parameter $u$ are 4.247 $\text{\AA}$ and 0.2048, respectively, which agree well with the experimental results \cite{TARASCON1981133}.
The calculation methods are given in Section A of the Supplemental Material (SM).
The point group of the structure is $O_h$, generated by the symmetries: $\hat I$, $\hat C_{4}^{x/y/z}$, $\hat C_2^{110/101/011}$, %$\hat C_2^{1\bar 10/10\bar 1/01\bar1}$
and $\hat C_{3}^{111}$, where $\hat C_m^{\bf{n}}$ is the $m$-fold rotation symmetry about the vector $\bf{n}$. Thus, one can define $\hat M_{\bf{n}}=\hat I \hat C_2^{\bf{n}}$, which is a mirror reflection about the plane defined by the normal vector $\bf{n}$.

\begin{figure*}[t]
%\captionstyle{centerlast}
%\includegraphics[clip,scale=0.5, angle=0]{wzj3}
\includegraphics[clip,scale=0.35, angle=0]{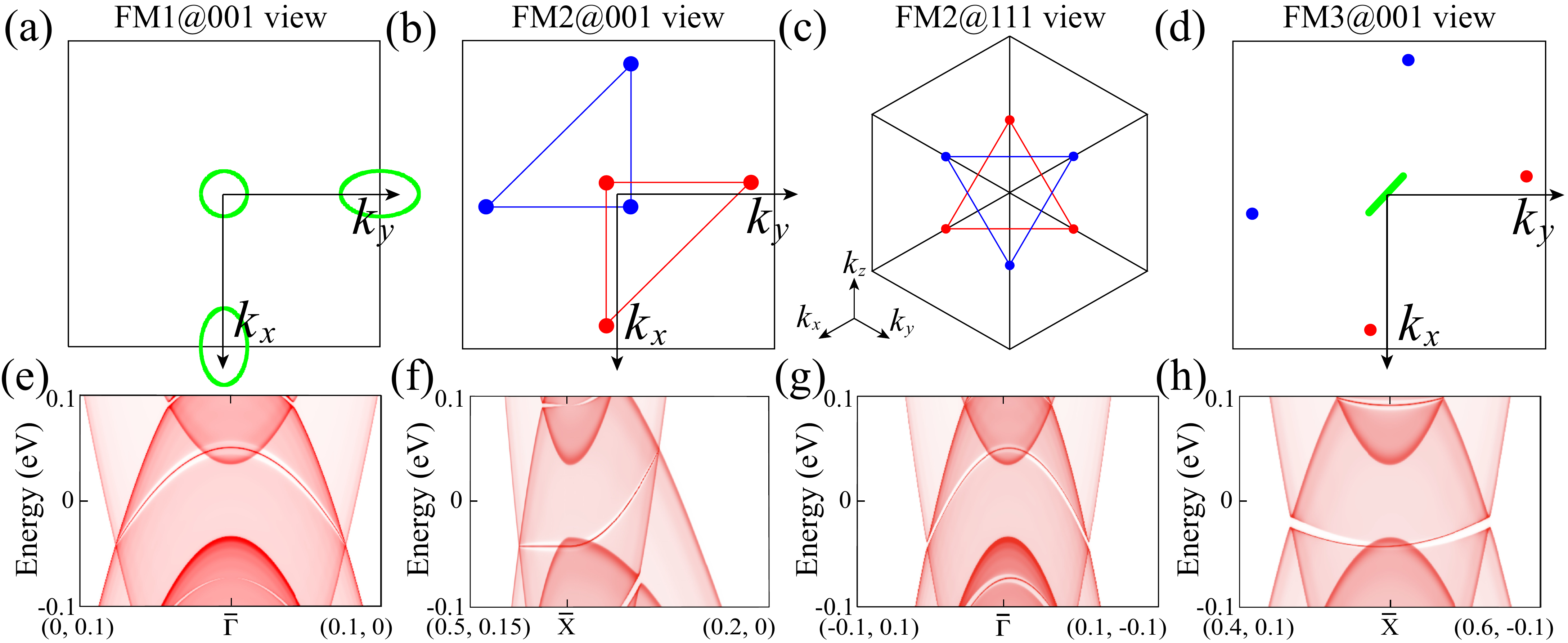}
 \caption{~$\boldsymbol{|}$ \textbf{Nodal lines, Weyl nodes and topological surface states in FM states of EuB$_6$.} %Without ambiguity, all the Weyl nodes are shifted to the first BZ in \textbf{(a)}-\textbf{(d)} .
\textbf{(a)} The 001-view of the FM1 state. \textbf{(b)} and \textbf{(c)} are the 001- and 111-view of the FM2 state. \textbf{(d)} The 001 view of the FM3 states. %Note that Z point is projected to the $\bar \Gamma$ point on the (001) surface.
%(e)-(h) show the (001) surface states for different FM phases and different Z point.
The green lines represent the nodal lines. The red and blue points represent the Weyl points with charge 1 and -1, respectively.
\textbf{(e)} and \textbf{(f)} are the surface states of the FM1 phase near the $\bar \Gamma$ and $\bar X$ points, respectively. \textbf{(g)} The surface states of the FM2 phase along (-0.1, 0.1)-$\bar \Gamma$-(0.1, -0.1). \textbf{(h)} The surface states of the FM3 phase along (0.4, 0.1)-$\bar X$-(0.6, -0.1).
The (001) surface states along $\bar M-\bar X-\bar \Gamma-\bar M$ can be found in Fig. S7 of the SM.
}
\label{fig:3}
\end{figure*}

\emph{PM state in high temperature.}
At high temperature, the magnetic order is absent. Here, an ``open core" treatment of Eu $4f$ electrons has been used to treat them as core states, %(far away from the Fermi energy),
which have negligible effect on the bands near E$_\text{F}$.
Considering the well-known underestimation of band gap within generalized gradient approximation (GGA), modified Becke-Johnson exchange (mBJ) calculation is performed to obtain accurate band structure. The corresponding results are presented in Fig. 2(c). It shows that PM EuB$_6$ is a semiconductor with a tiny direct gap ($\sim$ 20 meV) at three Z points [X, Y and Z points are symmetry-related (equivalent) by $\hat C_3^{111}$], which is consistent with the angle-resolved
photoemission spectroscopy (ARPES) \cite{eub6arpes}, low field Hall effect \cite{GUY19801055} and magnetoresistance measurements \cite{sullow2000metallization}.
Under the little group $(D_{4h})$ of the Z point, the valence band and conduction band near E$_\text{F}$ are labeled as $\Gamma_3^-$ and $\Gamma_3^+$, %(neglecting the spin degree of freedom)
respectively. These two bands with opposite parity play a critical role in further study of the magnetic states of EuB$_6$, as shown in Fig. 1.

\emph {FM states at low temperature.}
After FM transition, all the local moments of Eu $f^7$ configuration are aligning in the same direction. As the magnetism in EuB$_6$
is ``soft" \cite{sullow1998structure}, meaning that it can be easily tuned by temperature or an external magnetic field, we have first performed the calculations for the FM state without SOC.
%In this case, the two spin channels don't hybridize with each other and  the symmetry for each spin channel is $O_h$.%(without time-reversal symmetry $\cal T$).
The band structures for the spin-up (red solid lines) and spin-down (blue dashed lines) channels are shown in Fig. 2(d).
The highest valence band at Z point is mainly from an anti-bonding orbital formed by Eu $f$ and B $p$ states.
The spin-up valence orbital hybridizes strongly with the occupied $f$ states below E$_F$,
while the hybridization shift (level repulsion) in the spin-down channel is much weaker and of opposite sign, since the unoccupied $f$ bands are high above E$_F$ due to the on-site Coulomb repulsion.  As a result, we obtain an effective antiferromagnetic exchange coupling in the valence band.
However, the exchange coupling of the conduction band (mainly from Eu $d$ states) with local $f$ states is of ferromagnetic $f$-$d$ intra-atomic origin~\cite{kunevs2004kondo}. Therefore, the effective exchange splitting has opposite sign on the two low-energy bands. Because of this special exchange splitting,
band inversion happens at three Z points in the spin-up channel, while the normal band gap increases in the spin-down channel. It results in full spin polarization at E$_\text{F}$, consistent with previous theoretical calculations~\cite{massidda1996electronic, kunevs2004kondo}. Although there seems to be a conflict with some experimental results showing about $50\%$ spin-polarized state at E$_\text{F}$, the half-metallic state can be easily tuned into an incomplete spin-polarized state by light doping.  As most samples are electron- or hole-doped, the experimental results may sensitively depend on the chemical potential. The experimental observations of the metallic behavior in the FM states \cite{GUY19801055,nyhus1997} qualitatively support our conclusion of the inverted band structure. Since the correct electron correlation U is unknown and the internal parameter $u$ slightly varies from samples to samples, we have systematically investigated the phase diagram by computing the band gaps ({\it i.e.} $\eta=E_{\Gamma^+_3}-E_{\Gamma^-_3}$) in both spin-up and spin-down channels (see details in Section B of the SM). The obtained phase diagram shows that the band inversion of the spin-up channel survives in a large area of the parameter space, indicating much promise for finding the topological semimetal phases in EuB$_6$.
%In the presence of inversion symmetry, the band inversion between two bands with opposite parity represents an ideal toy model for the study of phase transitions between topological semimetal phases.

In the absence of SOC, the band inversion results in five intersecting nodal lines at the Z point protected by five mirror symmetries ($\hat M_z$, $\hat M_{x/y}$, $\hat M_{110/1\bar 1 0}$),
with three of which being shown schematically in the inset of Fig. 2(d). The similar situation happens at both X and Y points too. %which are related to the Z point by $\hat C_3^{111}$. %, as shown in Fig. \ref{structure}.
Then, we include SOC in the calculations and consider three FM states with different directions of magnetization (\emph{i.e.}, FM1$||[001]$, FM2$||[111]$ and FM3$||[110]$). The small energy difference between them indicates that the magnetic moment can be easily tuned (details in Section C of the SM).
%We note that SOC has two main effects on the FM states: (i) it mixes the two spin channel states and opens a hybridization gap; (ii) it changes the symmetry of the FM phases, which does effect the topology of the inverted bands.
After the consideration of SOC, a gap of about 0.01 eV will open along the nodal lines, as shown in Section C of the SM. However,
due to the band inversion resulting in $\chi=1$, the nodal lines cannot be fully gapped out, and
nodal lines or odd pairs of Weyl nodes are guaranteed around each Z point with an infinitesimal strength of SOC.
The exact situation strongly depends on the FM direction (symmetry), as will be shown below. %as shown in Fig. \ref{surf}.
%For instance, in [110] magnetism, a nodal line forms around the Z point, while a pair of Weyl nodes emerge around the X/Y point, as shown in Fig.~\ref{surf}(d).

\emph{Topology with different magnetic directions.}
In the case of FM1 with [001] magnetism, the symmetry reduces to the magnetic space group: $\{C_{4h} \oplus {\cal T}\hat M_{x} C_{4h}\}$, where the symmetry group $C_{4h}$ is generated by $\hat I$ and $\hat C_4^{z}$ (see details in Section D of the SM [SM D]).
Since $\hat{M}_{z}$ is still preserved and also belongs to the little group of three Z points, three nodal lines are expected, with one for each point. The calculated results are shown in Fig. 3(a). Consistently, two nodal rings (around X and Y, respectively) are found in the $k_z=0$ plane and one (around Z) in the $k_z=\pi$ plane.
Due to the strong anisotropy of the band dispersion, %in moment space,
the nodal line is oval-shaped around the X/Y point, while it's almost a circle around the Z point. %Based on the symmetry,
The $k\cdot p$ invariant  model Hamiltonian is constructed (more details in SM D) in the vicinity of each Z point, which gives exactly the same band crossing as obtained from the first-principles calculations.

When the magnetic field is aligned with the [111] direction (FM2), the magnetic space group becomes $\{C_{3i} \oplus {\cal T}\hat C_2^{1\bar 10} C_{3i}\}$, where %$\hat C_2^{1\bar 10}$ is a two-fold rotation symmetry about the $[1\bar 10]$ direction and
the symmetry group $C_{3i}$ is generated by $\hat I$ and $\hat{C}_3^{111}$.
At the Z (X, Y) point, $[{\cal T}\hat C_2^{1\bar 10 (01\bar 1, 10\bar 1)}]^2=1$ can stabilize Weyl nodes in the $k_x=k_y$ ($k_y=k_z$, $k_z=k_x$) plane.
We do find a pair of Weyl points for each Z point.
%(X, Y and Z points are related by $\hat C_3^{111}$), as shown in Fig. \ref{surf}(b) and (c).
The coordinates of two Weyl nodes near the Z point are found to be $\vec W_1=[\pm0.03978,\pm 0.03978,0.5\pm0.07854]$ (hereafter, the coordinates of $k$-points are given in units of [$\frac{2\pi}{a},\frac{2\pi}{a},\frac{2\pi}{a}$]).
The corresponding model Hamiltonians are derived in SM C and yield the consistent results.
The other two pairs around the X and Y points are obtained by $\hat C_3^{111}$ (See the exactly coordinates in Section E of the SM [SM E]). As a result, the (111) plane through X, Y and Z points, denoted by a dashed triangle in Fig. S4(a), has a nontrivial Chern number $\text{C}=3$ (see details in SM E), %(the Wilsonloop flow is given in SM E),
which is crucial to realize large-Chern-number QAHE in its [111]-oriented quantum-well structure.

When the magnetic field is aligned with the [110]-direction, the symmetry reduces to the magnetic space
group: $\{C_{2h} \oplus {\cal T}\hat C_2^{z} C_{2h}\}$, where the symmetry group $C_{2h}$ is
generated by $\hat I $ and $\hat C_2^{110}$.
Interestingly, Z point is not equivalent to X or Y point any more. Although $\hat{M}_{z}$ symmetry is broken by the [110] magnetism, $\hat{M}_{110}$ symmetry restores. As shown in Fig. 3(d), one nodal line (shown as the green line) circled around Z point in the $k_x+k_y=0$ plane still survives due to the protection of $\hat{M}_{110}$ symmetry.
However, at X or Y point, there is no any mirror symmetry. A pair of Weyl nodes are found in the $k_z=0$ plane, which are stabilized by the combined anti-unitary symmetry with the relation $[{\cal T}C_2^{z}]^2=1$ \cite{Wang2016MoTe2}. The coordinates of the Weyl nodes near X point are found to be $\vec W_2=[0.5\pm0.06204$, $\pm 0.06287$, $0]$,  and the Weyl nodes near Y point can be obtained by $\hat C_2^{110}$ symmetry. %(see SM E).
As expected, the (110) plane through both X and Y points has a nonzero Chern number $\text{C}=2$ (see details in SM E).
%All the first-principles results are in good agreement with the results from the $k\cdot p$ models in SM D.

\begin{figure}[t]
 %\captionstyle{centerlast}
%\includegraphics[clip,scale=0.25, angle=0]{wzj4}
\includegraphics[clip,scale=0.24, angle=0]{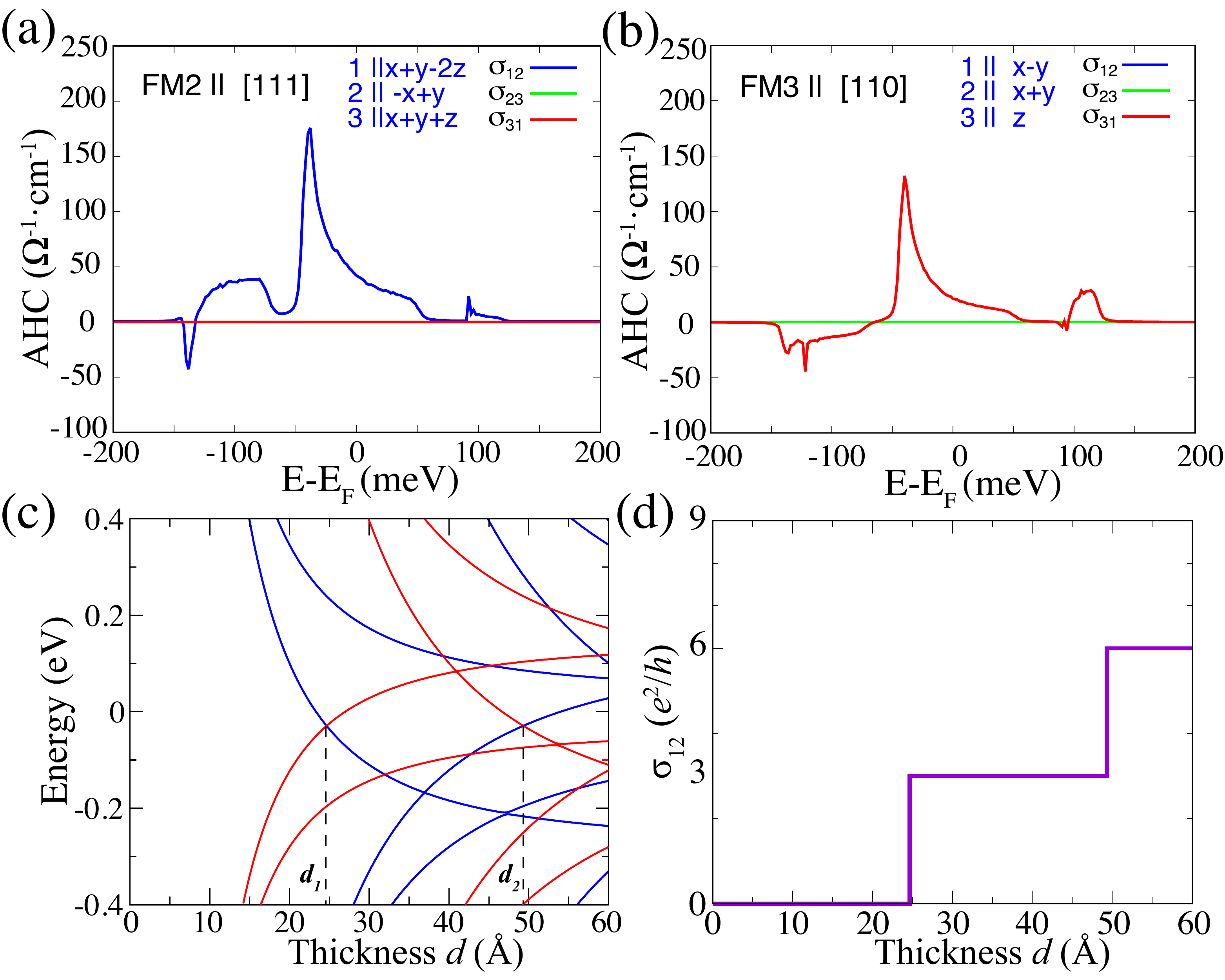}
\caption{~$\boldsymbol{|}$ \textbf{AHC and quantized Hall conductance.} \textbf{(a)} and \textbf{(b)} Energy dependent AHC of FM2 EuB$_6$ and FM3 EuB$_6$, respectively.
\textbf{(c)} The evolution of the band energies at $Z$ point as a function of the film thickness ($d$). \textbf{(d)}
The quantized Hall conductance as a function of the film thickness ($d$).
}
\label{fig:4}
\end{figure}

\emph{Topological surface states and AHC.}
The nontrivial band structures suggest the existence of topological nontrivial surface states, which are calculated
with the Green's function methodology \cite{marzari2012maximally}. The results are shown in Fig. 3. For FM1 EuB$_6$ %the FM1 phase of EuB$_6$
with a nodal line circled around each Z point, drumhead surface states are obtained within the energy gap at $\bar{\Gamma}$ (Fig. 3(e)) and $\bar{\text{X}}$ (Fig. 3(f)). Interestingly, the surface states are almost flat along $\bar{\text{X}}\bar {\text{M}}$.
For the WSM phase in FM2 EuB$_6$, %of the FM2 phase,
a chiral surface mode is obtained along (-0.1, 0.1)-(0.1, -0.1), as shown in Fig. 3(g). It can be regraded as the edge state of the Chern-insulator plane (spanned by $[1\bar 1 0]$ and [001]), because the plane separates two opposite Weyl nodes near the Z point.
Similarly, in the FM3 phase, one chiral surface state near $\bar{\text{X}}$ is obtained along (0.4, 0.1)-(0.6, -0.1), as shown in Fig.~3(h).
The nontrivial surface states should be observable in future experiments using ARPES or scanning tunnel microscope.
In addition, anomalous Hall effect is another exotic consequence of magnetic WSMs.
%The magnitude of AHC can be easily evaluated by $\sigma_{xy}=\frac{e^2}{h}\frac{\Delta k_z^W}{2\pi}$,  where $\Delta k_z^W$ is the distance between Weyl nodes projected onto the $z$ axis. %As thus, the evaluated values are $499.58$ and $322.27$ $(\Omega\cdot cm)^{-1}$ for the FM2 and FM3 phases, respectively.
The calculated AHC are shown in Figs.~4(a) and (b) (the coordinates are redefined, with a new axis parallel to the magnetic direction),
%As it's spin-polarized, only the component of AHC in the magnetic direction is nonzero.
while the AHC in FM1 phase is almost zero near E$_F$ (not shown in Fig.~4).
%The calculated values are relatively smaller than the evaluated values. This is because: i) the band structure is complicated due to the anisotropy. Namely, the energy bands vary in a large energy range; ii) the hybridizing gap is small due to the weak SOC strength.
In fact, the anomalous Hall effect has been observed in the bulk magnetic EuB$_6$ \cite{wigger2004}.
%The small density of carries \cite{sullow2000magnetotransport} and substantial AHC could lead to a large Hall angle, which can be also revealed in the future measurements.

\emph{QAHE in quantum-well structures.}
Magnetic WSM can be viewed as a stack of two-dimensional Chern insulator with strong coupling in the stacking direction.
By considering its quantum-well structure, quantized Hall effect can be achieved due to the confinement effect, without an additional magnetic field, which is also known as QAHE. Based on the effective $k\cdot p$ models in SM D, the Hamiltonians for
the [111]-oriented quantum wells of FM2 EuB$_6$ have been constructed (see details in section F of the SM). The evolution of the low-energy
 sub-bands %(labeled as $\ket{H_n}$ and $\ket{E_n}$ for hole and electron subbands, respectively)
 at Z point as a function of the film thickness ($d$) is calculated and shown in Fig. 4(c).
When the film is very thin, the band inversion in the bulk FM2 EuB$_6$ is removed and
the film is a trivial insulator. After the first critical thickness (about $24$ \AA), band inversion happens between a
hole sub-band and a electron sub-band with even (red lines) and odd (blue lines) parity, respectively.
Consequently, it leads to a jump in the Chern number or the Hall coefficient $\sigma_{12}$ \cite{Liu2008}, as shown in Fig.~4(d).
We find subsequent jumps of $\sigma_{12}$ in unit of $3e^2/h$, because band inversion happens at the
three Z points related by $\hat C_{3}^{111}$.
%The Wannier charge center calculation shows that the film becomes a QAHI with Chern number $C=3$.  If the film becomes more thick, larger number QAHI with transverse Hall conductance $\sigma_{xy}=\frac{Ce^2}{h}$ can be realized, as shown in Fig. \ref{ahc}(d).
As the thin film of EuB$_6$ has been grown successfully \cite{bachmann1970spin}, the realization of large-Chern-number QAHE in EuB$_6$ is experimentally realizable.

\emph{Conclusion.}
In summary, we have studied topological phases in FM EuB$_6$ adopting first-principles calculations and effective models. In the PM phase, two bands near E$_\text{F}$ have different parity eigenvalues at three Z points. The effective magnetic exchange splitting has opposite effect on the two bands, leading to band inversion in the spin-up channel. The calculated phase diagram shows that the band inversion between two bands with opposite parity survives in a large region. This kind of band inversion guarantees the existence of either nodal lines or odd pairs of Weyl nodes.
%As expected, there are three nodal lines in the FM1 phase, three pairs of Weyl nodes in the FM2 phase, and one nodal line and two pairs of Weyl nodes in the FM3 phase.
Generally speaking, even though the magnetic moment is tuned in an arbitrary direction (breaking all mirror symmetries), odd pairs of Weyl nodes are still guaranteed around each Z point. Topological surface states and AHC are obtained. % and call for more experiments.
We find that the AHC is sensitive to the magnetic order, which can be further measured in experiments.
In the [111]-oriented quantum-well structure of FM2 EuB$_6$, large-Chern-number QAHE can be achieved.

~\\
\emph{Acknowledgements.}---
S. N. and F. P. are supported by Volkswagen of America and the Affiliates Program of the Nanoscale Protoyping Laboratory .
   Z. W. is supported by the National Thousand-Young-Talents Program, the CAS Pioneer Hundred Talents Program, and the the National Natural Science Foundation of China.
   X.D. and H.M.W. are supported by the Ministry of Science and Technology of China (grant no. 2016YFA0300600) and the K.C. Wong Education Foundation (grant no. GJTD-2018-01).
   X.D. acknowledges financial support from the Hong Kong Research Grants Council (project no. GRF16300918).

\bibliography{EuB6}

\ \\

\clearpage

\begin{widetext}
\newpage
\beginsupplement{}
\section*{SUPPLEMENTAL  MATERIAL}

\subsection{A. Calculation methods}
The calculations were performed using the full-potential linearized-augmented plane-waves (FP-LAPW) method as implemented in the WIEN2K package \cite{blaha2002wien2k}.
The exchange-correlation in the parametrization of Perdew, Burke and Ernzerhof within GGA was applied in the calculations \cite{pbe}. SOC was included as a second variational step self-consistently. The radii of the muffin-tin sphere $R_{MT}$ were 2.5 and 1.55 bohrs for Eu and B, respectively. The truncation of the modulus of the reciprocal lattice vector $K_{max}$ was set to $R_{MT}K_{max}=7$. The $k$-point sampling grid of the BZ was $20\times 20\times 20$.  The geometry optimization was carried out until the maximal force on each ion becomes less than 0.01 eV.\AA$^{-1}$. The optimized lattice constant $a$ and internal parameter $u$ are 4.247 $\text{\AA}$ and 0.2048, respectively, which agree well with the experimental results \cite{TARASCON1981133}. In order to simulate the high-temperature PM state, an $``\text{open core}"$ treatment of $4f$ electrons was used in effect regarding these as core electrons \cite{tran2009accurate}. In this approach, the rare earth ion Eu$^{2+}$ behaves like nonmagnetic alkali earth ion (\emph{e.g.,} Ca$^{2+}$).
Considering the correlation effect of the $f$-electrons, the GGA+Hubbard-U (GGA+U) method was adopted, and the exchange parameter $U=8$ eV was applied to the Eu $4f$ states for the calculations in FM phases \cite{SICLDAU}.
Maximally localized Wannier functions  were constructed to calculate the surface states of the semi-infinite samples using an iterative method \cite{marzari2012maximally, wu2017wanniertools}.

\subsection{B. Phase diagrams with varying U and u}
By studying the band structures of FM EuB$_6$, one can find that the internal parameter $u$ and Hubbard U have influences on the band inversion at the Z points ($u$ is defined as a fractional parameter related to the lattice constant $a$). Without loss of generality, the impact of $u$ and U on the band inversion has been
systematically investigated with the fixed lattice constant $a=4.247$ $\text{\AA}$. SOC is ignored in these calculations. The band gaps (defined as $\eta =E_{\Gamma_{3}^+} -E_{\Gamma_{3}^-}$) in the spin-up and spin-down channels are presented by color in Fig.~\ref{phase}(a) and \ref{phase}(b), respectively.
In Fig.~\ref{phase}(a), one can find that, although the depth of the band inversion ($|\eta|$) at Z point decreases as increasing the internal parameter $u$ or Hubbard U, the band inversion happens in the whole parameter region for the spin-up channel, indicating the robustness of the band inversion in FM EuB$_6$.
In Fig.~\ref{phase}(b), we find U barely changes the band gap $\eta$ in the spin-down channel. When $u$ is very small (such as 0.202), the band inversion happens also in the spin-down channel. However, the increase of $u$ can remove it.

In conclusion, the situation of the band inversion in the spin-up channel and no band inversion in the spin-down channel can survive in a large range of the phase diagrams, which is what we consider in the main text. The points of the parameters used in the main text are indicated by yellow stars.
%small $u$, while only spin-up channel has band inversion for the samples with large $u$ (such as 0.207).
%The distribution of Weyl points for both cases will be discussed below.

\begin{figure}[h]
\includegraphics[clip,scale=0.40, angle=0]{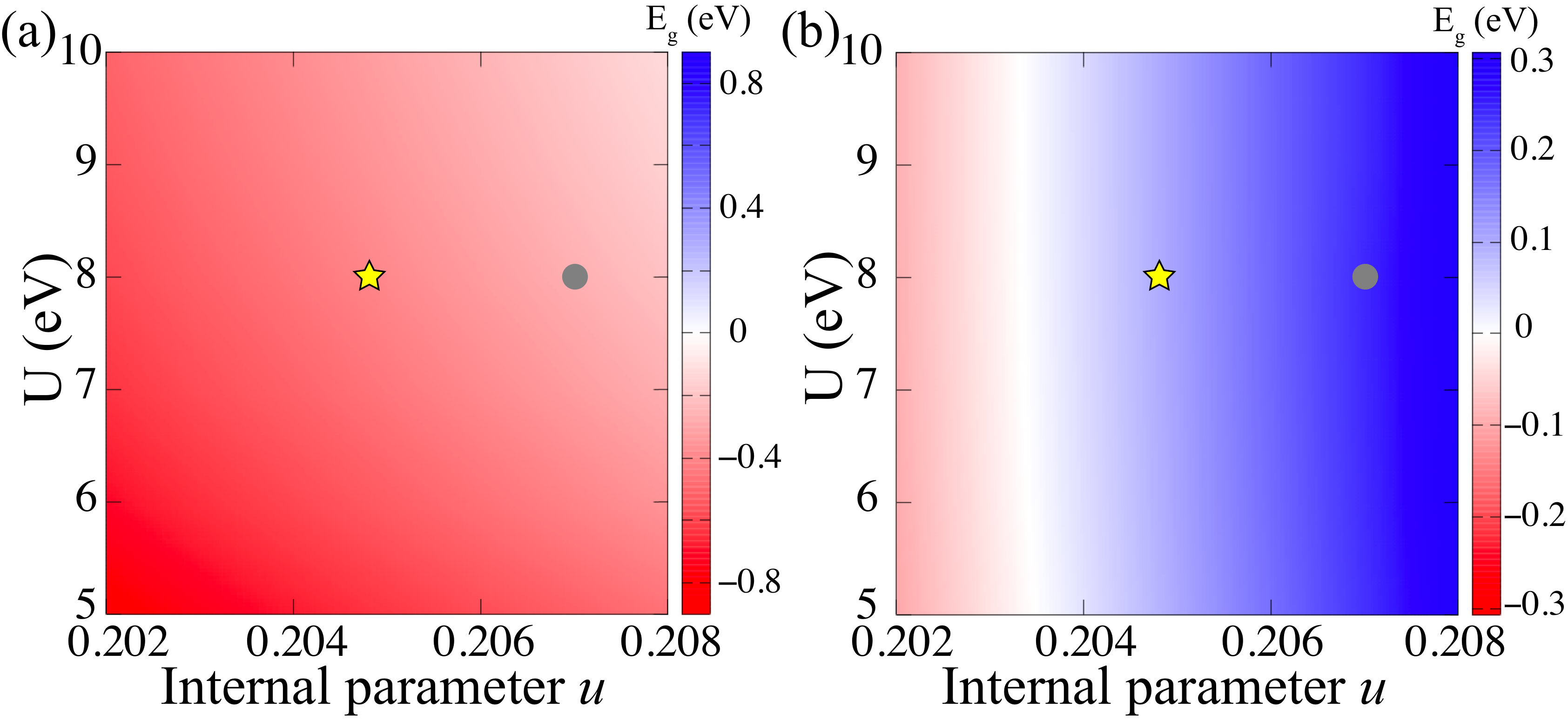}
\caption{ The band gaps (defined as $\eta =E_{\Gamma_{3}^+} -E_{\Gamma_{3}^-}$) at the Z point for the spin-up \textbf{(a)} and spin-down \textbf{(b)} channels of FM EuB$_6$ as a function of the internal parameter $u$ and Hubbard U. The position of the optimized parameters is indicated by yellow stars. The position of FM2 EuB$_\text{6}$ with nine pairs of Weyl nodes is indicated by gray circles.}
\label{phase}
\end{figure}

\begin{table}[t]
\begin{center}
\caption{The total energies of difference FM phases calculated by WIEN2k. The total energy of the FM2 state is chosen as the reference energy to get the energy difference. The energy is given in units of meV/f.u..}%(f.u.$\equiv$ formula unit).}
\label{totene}
\begin{tabular}{p{3cm}|p{3cm}|p{3cm}|p{3cm}}
\hline
\hline
 Phase ~~~~~~~~~~~~~~~~~~~~~~~~       & FM1 & FM2 & FM3 \\
\hline
Total energy  & -299371141.250 & -299371163.976 & -299371163.973 \\
\hline
Energy difference & 22.726 &  0 & ~0.003\\
\hline
\hline
\end{tabular}
\end{center}
\end{table}

\begin{figure*}[t]
 %\captionstyle{centerlast}
\includegraphics[clip,scale=0.29, angle=0]{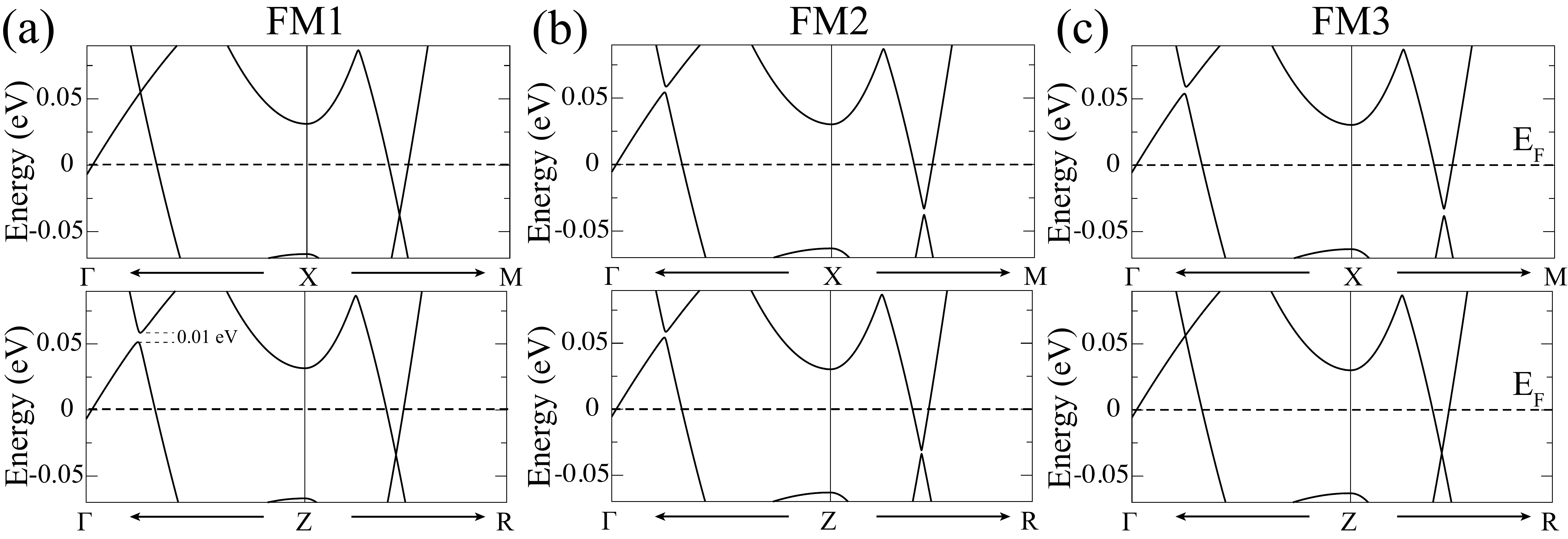}
\caption{GGA+U+SOC band structures around X and Z points for the FM1 \textbf{(a)}, FM2 \textbf{(b)} and FM3 \textbf{(c)} states of EuB$_6$, respectively.
}
\label{bands}
\end{figure*}

\subsection{C. Total energies and band structures for different FM states}
In order to confirm the ground FM state of EuB$_6$, we have performed the GGA+U+SOC calculations for the three FM states, \emph{i.e.}, FM1$||[001]$, FM2$||[111]$ and FM3$||[110]$, as shown in Table \ref{totene}.
%The results show that the total energy of the FM3 state is about 22.723 meV/f.u. lower than that of the FM1 state, and the FM2 state is the ground state, further lowering the total energy about 0.003 meV/f.u. than the FM3 state.
The results show that the FM2 state is the ground state, which has the lowest total energy. Relative to that of the FM2 state, total energies of the FM1 and FM3 states are 22.726 meV and 0.003 meV higher, respectively.
Our results are consistent with the experimental measurements \cite{sullow1998structure, brooks2004magnetic}. The small energy difference between the FM3 state and FM2 state (at the order of $10^{-3}$ meV), beyond the accuracy of first-principles calculations, indicates the possibility to realize FM3 state in experiments. The different long-range FM orders selectively break specific symmetries and lead to different bulk topological properties, as discussed in the main text.

By using GGA+U+SOC, the detailed band structures of the three FM states are also calculated around the X and Z points, as shown in Fig. \ref{bands}. Because of the breaking of the four mirror symmetries
({\it i.e.} $\hat M_{x/y}$, $\hat M_{110/1\bar 1 0}$), and the presence of $\hat M_z$ symmetry in the FM1 state,
a band gap of 0.01 eV is opened at the band crossing on the line $\Gamma-Z$, while the band crossings on the lines $Z-R$, $\Gamma-X$, and $X-M$ are protected by $\hat M_z$, as shown in Fig. \ref{bands}(a).
For the FM2 state, all the band crossings are gapped out due to the breaking of all mirror symmetries, as shown in Fig. \ref{bands}(b).
For the FM3 state, the band crossings on the lines $\Gamma-X$ and $X-M$ are gapped out, while the band crossings on the lines $\Gamma-Z$ and $Z-R$ survive because of the $\hat M_{110}$ symmetry protection, as shown in Fig. \ref{bands}(c).

%When the FM transition happens, the Eu$^{+2}$ $f^7$ ions with collinear magnetic structure have completely
%occupied spin-up $f$ states and unoccupied spin-down $f$ states.  Both the theoretical calculations and experimental
%results show the semi-metallic property of FM EuB$_6$. However, the half-metallic picture predicted by theoretical
%calculations is in contradiction with the incomplete spin polarization observed by both Andreev reflection
%spectroscopy \cite{zhang2008spin} and resonant inelastic x-ray scattering \cite{kim2013}. Here, the band
%structures of FM EuB$_6$ are studied by GGA+U method.

%\newpage
\subsection{D. Effective $k\cdot p$ models}
At high temperature, the PM phase of EuB$_6$ has the space group: $\{O_h \oplus {\cal T}O_h\}$.
The crystal point group $O_h$ contains a total of 48 symmetry operations, including identity {$\hat{E}$}, inversion symmetry ($\hat{I}$), nine mirror symmetries (such as $\hat{M}_x$, $\hat{M}_y$ and $\hat{M}_z$), nine twofold rotational symmetries (such as $\hat{C}_2^{z}$), sixteen threefold rotational symmetries (for example $\hat{C}_3^{111}$), and twelve fourfold rotational symmetries (for instance $\hat{C}_4^{z}$).

\begin{table}[b]
\begin{center}
\caption{The parameters for $\epsilon_\bk$, $M_{\bk}$, $d_\bk$ and $f_{\bk}$.}
\label{para0}
\begin{tabular}{ccccccc}
\hline
\hline
\multicolumn{3}{c}{Units}& & [eV] & [eV$\cdot$\AA$^2$] & [eV$\cdot$\AA$^2$] \\

\hline
$\epsilon_0$ & $\epsilon_1$ & $\epsilon_2$&:& -0.032425 & 3.050842 &1.142637 \\
\hline
$M_0$ & $M_1$ & $M_2$ &:& -0.089685 & 5.153294 & 15.768395 \\
\hline
$d_0$ & $d_1$ & $d_2$ &:& -0.156770 & 0.479908 & 1.828220\\
\hline
$f_0$ & $f_1$ & $f_2$ &:& 0.090810 & -0.457055 & -0.457055\\
\hline
\hline
\end{tabular}
\end{center}
\end{table}

\subsubsection{1. Symmetries and the $k\cdot p$ models for the FM state without SOC}
When SOC is ignored, the spin and spatial rotation can be treated separately. The FM phase of EuB$_6$ without SOC has the symmetry $\{O_h\}$.
To capture the band structures in the phase transition, we first derive a 4$\times$4 model Hamiltonian to describe the low-energy physics of EuB$_6$ in the FM state without SOC. %, which implies it's SU(2) symmetric in the spin degree of freedom.
As mentioned in the main text, the bands near the Fermi level belong to $\Gamma^{3+}$ and $\Gamma^{3-}$.
The prototype basis of $\Gamma_3^+$ is the $\vert d_{x^2-y^2} \rangle$ state, and that of $\Gamma_3^-$ is the $\vert f_{xyz} \rangle$ state.
Note that these real-orbital labels do  not have to indicate the main orbital component of the wave functions.
Under the basis $\{ [\ket\up,\ket\dn]\otimes[\vert d_{y^2-z^2} \rangle,\vert f_{xyz}\rangle]\}$, the magnetic effective model at Z point (without SOC) is obtained as follows:
\begin{eqnarray}
&H^{[\theta,\phi]}_Z(\bk)=\epsilon_\bk  s_0 \tau_0 + M_\bk    s_0 \tau_z  + s^{[\theta,\phi]} \left(\begin{array}{cc}
      d_\bk &0 \\
      0 &f_\bk \end{array}\right), \\
%\begin{equation}
%\begin{split}
&\text{with }\bk\equiv (k_x,k_y,k_z)=(K_x,K_y,K_z)-\vec K^Z;~s^{[\theta,\phi]}\equiv
\left(
\begin{array}{cc}
 \cos \left(\frac{\theta }{2}\right) & -\sin \left(\frac{\theta }{2}\right) \\
 e^{i \phi } \sin \left(\frac{\theta }{2}\right) & e^{i \phi } \cos \left(\frac{\theta }{2}\right) \\
\end{array}
\right). \notag
\end{eqnarray}
Note that $\bk\equiv(k_x,k_y,k_z)$ is the displacement from the Z point.
($K_x,K_y,K_z$) are the absolute values of the $k$-point in the Cartesian coordinate, and $\vec K^Z$ is (0, 0, $\frac{\pi}{c}$).
%The parameters of $\epsilon_\bk$, $M_{\bk}$, $d_\bk$ and $f_{\bk}$ in Eq. (\ref{emdf}) below are given in Table.~\ref{para0}.
$\epsilon_\bk$, $M_{\bk}$, $d_\bk$ and $f_{\bk}$ take the following forms
\begin{equation}
\begin{split}
\epsilon_\bk= &\epsilon_0+\epsilon_1 k_z^2+\epsilon_2(k_x^2+k_y^2);\\
M_\bk= &M_0+M_1 k_z^2+M_2(k_x^2+k_y^2) ;\\
d_\bk=&d_0+d_1 k_z^2+d_2(k_x^2+k_y^2) ;\\
f_\bk=&f_0+f_1 k_z^2+f_2(k_x^2+k_y^2) .\\
\end{split}
\label{emdf}
\end{equation}
where $s_i (\sigma_i)$ are Pauli matrices, acting on the spin (orbital) component. The parameters ($\epsilon_i$, $M_i$, $d_i$ and $f_i$ with $i=0$, 1, 2) are given in Table.~\ref{para0}.
The $k\cdot p$ Hamiltonians for X and Y points (with $\bk$ defined as the displacements from the X and Y points, respectively) are given as,
%\begin{equation}
%\begin{split}
\begin{eqnarray}
&H^{[\theta,\phi]}_X(k_x,k_y,k_z)=H^{[\theta,\phi]}_Z(k_y,k_z,k_x) \\
&H^{[\theta,\phi]}_Y(k_x,k_y,k_z)=H^{[\theta,\phi]}_Z(k_z,k_x,k_y)
\end{eqnarray}
%\end{split}
%\end{equation}

\begin{figure}[h]
\includegraphics[clip,scale=0.15, angle=0]{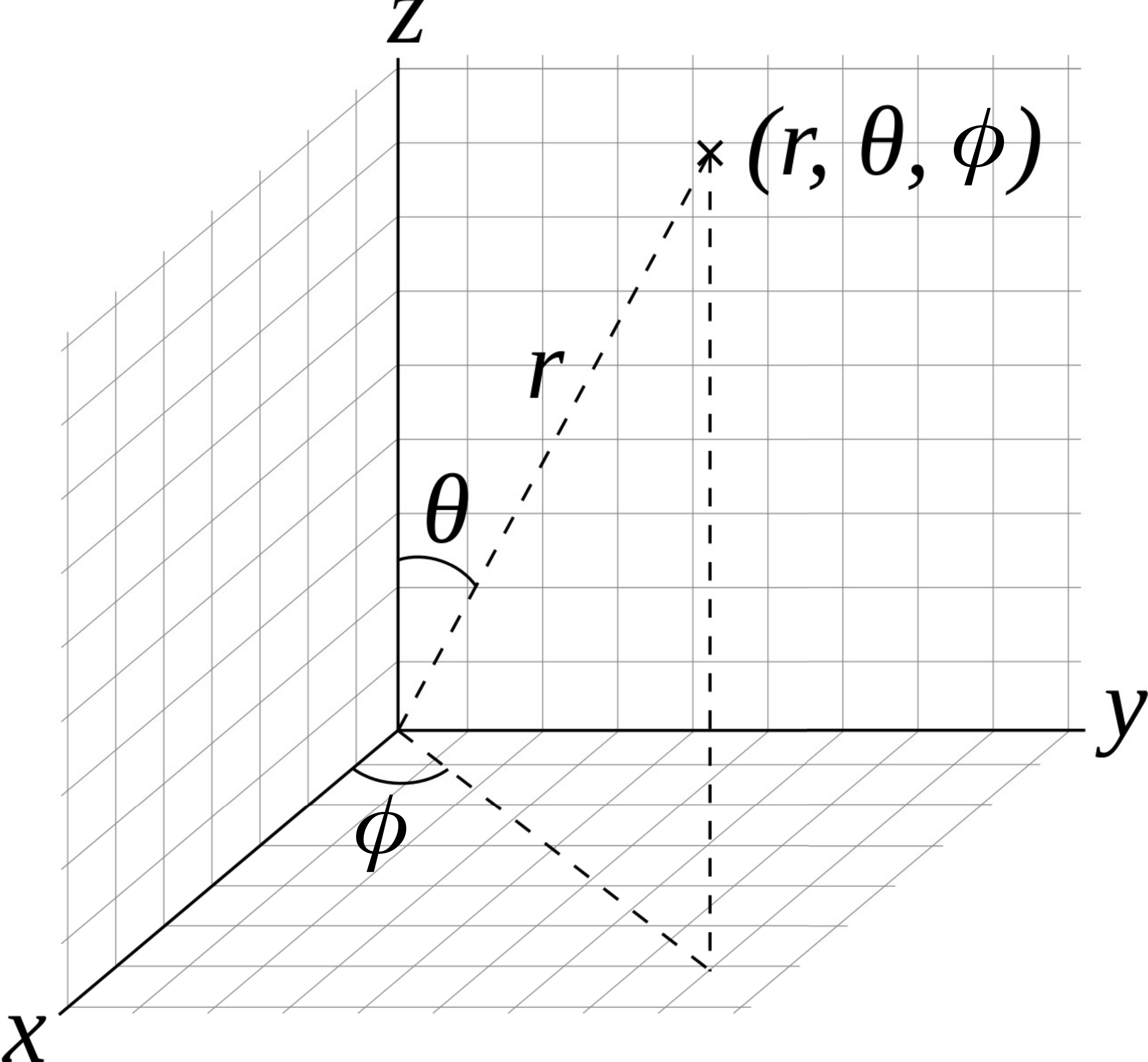}
\caption{ Spherical coordinates ($r,~\theta,~\phi$) as commonly used in physics (ISO convention): radial distance $r$, polar angle ($\theta$), and azimuthal angle ($\phi$). The polar angle $\theta$ and azimuthal angle $\phi$ also define the direction of a vector.
}
\label{3Dthephi}
\end{figure}

The polar angle $\theta$ and azimuthal angle $\phi$ (shown in Fig.~\ref{3Dthephi}) indicate the direction of the magnetism ({\it e.g.} [001], [111], [110]). Therefore, the superscript $[\theta,\phi]$ can be also changed to the direction, directly.
The values for some typical directions are given in Table.~\ref{thephi}.

\begin{table}[tb]
\caption{The values of the polar angle ($\theta$) and azimuthal angle ($\phi$) for some typical directions.}
\begin{tabular}{c c c c}
\hline
\hline
 Directions & $\theta$  & $\phi$ &$s^{[\theta,\phi]}$ \\
\hline
[001] & 0 & \emph{Arbitrary} & $s_z$ \\
\hline
[110] & $\pi$/2 & $\pi$/4 & $(s_x+s_y)/\sqrt{2}$ \\
\hline
[111] & $arccos[\frac{1}{\sqrt{3}}]$& $\pi$/4 & $(s_x+s_y+s_z)/\sqrt{3}$\\
      & ($\sim 0.304086724\pi$) & & \\
\hline
\hline
\end{tabular}
\label{thephi}
\end{table}

\subsubsection{2. $k\cdot p$ models in FM1 phase with SOC}
After the consideration of SOC, the spin space and orbital space are coupled togegher. Magnetic space group is used to describe the symmetry of a magnetic system. When moment is aligned along [001] direction, the magnetic space group is
$\{C_{4h} \oplus {\cal T}\hat M_{x} C_{4h}\}$, containing $\hat{E}$, $\hat{I}$, $\hat{C}_4^{z}$,
$\hat{C}_4^{00\bar{1}}$, $\hat{C}_2^{z}$, $\hat{M}_{z}$, $\hat{I} \hat{C}_4^{z}$, $\hat{I}\hat{C}_4^{00\bar{1}}$,
${\cal T}\hat{C}_2^{y}$, ${\cal T}\hat{C}_2^{1\bar{1}0}$, ${\cal T}\hat{C}_2^{110}$,
${\cal T}\hat{C}_2^{x}$, ${\cal T}\hat{M}_{y}$, ${\cal T}\hat{M}_{1\bar{1}0}$, ${\cal T}\hat{M}_{110}$,
${\cal T}\hat{M}_{x}$.

The symmetries of the Z point with [001] magnetism contain
$\hat I$, $\hat C_4^{z}$, ${\cal T}\hat C_2^{x/y}$, ${\cal T}\hat C_2^{110/1\bar 10}$.
The model Hamiltonian for the point is obtained as,
\begin{equation}
\begin{split}
H^{FM1}_Z(\bk)=&H^{[001]}_Z(\bk)+ (Ak_xs_x+Ak_ys_y+Ck_zs_z)\tau_x \\
 \text{with } & A=C=0.04~\text{eV}\cdot\text{\AA};
\end{split}
\end{equation}

The symmetries of the X point with [001] magnetism contain
$\hat I$, $\hat C_2^{z}$, ${\cal T}\hat C_2^{x}$, ${\cal T}\hat C_2^{y}$.
The model Hamiltonian for the point is obtained as,
\begin{equation}
\begin{split}
H^{FM1}_X(\bk)=&H^{[001]}_X(\bk)+ (Ak_xs_x+Bk_ys_y+Ck_zs_z)\tau_x \\
 \text{with } & A=B=C=0.04~\text{eV}\cdot\text{\AA};
\end{split}
\end{equation}

The symmetries of the Y point with [001] magnetism contain
$\hat I,\hat C_2^{z}, {\cal T}\hat C_2^{x}, {\cal T}\hat C_2^{y}$.
The model Hamiltonian for the point is obtained as,
\begin{equation}
\begin{split}
H^{FM1}_Y(\bk)=&H^{[001]}_Y(\bk)+ (Ak_xs_x+Bk_ys_y+Ck_zs_z)\tau_x \\
 \text{with } & A=B=C=0.04~\text{eV}\cdot\text{\AA};
\end{split}
\end{equation}

\subsubsection{3. $k\cdot p$ models in FM2 phase with SOC}
When moment is aligned along [111] direction, the magnetic space group is
$\{C_{3i} \oplus {\cal T}\hat C_2^{1\bar 10} C_{3i}\}$, containing
$\hat{E}$, $\hat{I}$, $\hat{C}_3^{111}$, $\hat{C}_3^{\bar{1}\bar{1}\bar{1}}$,
$\hat{I}\hat{C}_3^{111}$, $\hat{I}\hat{C}_3^{\bar{1}\bar{1}\bar{1}}$,
${\cal T}\hat{C}_2^{01\bar{1}}$, ${\cal T}\hat{C}_2^{1\bar{1}0}$, ${\cal T}\hat{C}_2^{\bar{1}01}$
${\cal T}\hat{M}_{01\bar{1}}$, ${\cal T}\hat{M}_{1\bar{1}0}$, and ${\cal T}\hat{M}_{\bar{1}01}$.

The symmetries of the Z point with [111] magnetism contain
$\hat I$ and ${\cal T}\hat C_2^{1\bar 10}$.
The model Hamiltonian for the point is obtained as,
\begin{equation}
\begin{split}
H^{FM2}_Z(\bk)=&H^{[111]}_Z(\bk)+ (Ak_xs_x+Bk_ys_y+Ck_zs_z)\tau_x \\
&+D(k_x-k_y)s_z\tau_y \\
 \text{with } &A=B=-C=D=0.04~\text{eV}\cdot\text{\AA};
\end{split}
\end{equation}

The symmetries of the X point with [111] magnetism contain
$\hat I$ and ${\cal T}\hat C_2^{01\bar1}$.
The model Hamiltonian for the point is obtained as,
\begin{equation}
\begin{split}
H^{FM2}_X(\bk)=&H^{[111]}_X(\bk)+ (Ak_ys_y+Bk_zs_z+Ck_xs_x)\tau_x \\
&+D(k_y-k_z)s_x\tau_y \\
 \text{with } &A=B=-C=D=0.04~\text{eV}\cdot\text{\AA};
\end{split}
\end{equation}

The symmetries of the Y point with [111] magnetism contain
$\hat I$ and ${\cal T}\hat C_2^{10\bar 1}$.
The model Hamiltonian for the point is obtained as,
\begin{equation}
\begin{split}
H^{FM2}_Y(\bk)=&H^{[111]}_Y(\bk)+ (Ak_zs_z+Bk_xs_x+Ck_ys_y)\tau_x \\
&+D(k_z-k_x)s_y\tau_y \\
 \text{with } &A=B=-C=D=0.04~\text{eV}\cdot\text{\AA};
\end{split}
\end{equation}

\subsubsection{4. $k\cdot p$ models in FM3 phase with SOC}
When moment is aligned along [110] direction, the magnetic space group is
$\{C_{2h} \oplus {\cal T}\hat C_2^{z} C_{2h}\}$, containing
$\hat{E}$, $\hat{I}$, $\hat{C}_2^{110}$, $\hat{M}_{110}$, ${\cal T}\hat{C}_2^{z}$,
${\cal T}\hat{C}_2^{1\bar{1}0}$, ${\cal T}\hat{M}_{z}$, and ${\cal T}\hat{M}_{1\bar{1}0}$.

The symmetries of the Z point with [110] magnetism contain
$\hat I,\hat C_2^{110}, {\cal T}\hat C_2^{z}, {\cal T}\hat C_2^{1\bar 10}$.
The model Hamiltonian for the point is obtained as,
\begin{equation}
\begin{split}
H^{FM3}_Z(\bk)=&H^{[110]}_Z(\bk)+ (Ak_xs_x+Bk_ys_y+Ck_zs_z)\tau_x \\
 \text{with } & A=B=C=0.04~\text{eV}\cdot\text{\AA}.
\end{split}
\end{equation}

The symmetries of the X point with [110] magnetism contain
$\hat I$ and $\hat C_2^{z}$.
The model Hamiltonian for the point is obtained as,
\begin{equation}
\begin{split}
H^{FM3}_X(\bk)=&H^{[110]}_X(\bk)+ (Ak_xs_x+Bk_ys_y+Ck_zs_z)\tau_x \\
               &+Dk_zs_x\tau_y \\
 \text{with } & A=-B=C=D=0.04~\text{eV}\cdot\text{\AA}.
\end{split}
\end{equation}

The symmetries of the Y point with [110] magnetism contain
$\hat I$ and $\hat C_2^{z}$.
The model Hamiltonian for the point is obtained as,
\begin{equation}
\begin{split}
H^{FM3}_Y(\bk)=&H^{[110]}_Y(\bk)+ (Ak_xs_x+Bk_ys_y+Ck_zs_z)\tau_x \\
               &+Dk_zs_y\tau_y \\
 \text{with } & -A=B=C=-D=0.04~\text{eV}\cdot\text{\AA}.
\end{split}
\end{equation}

\begin{figure}[h!]
\includegraphics[clip,scale=0.25]{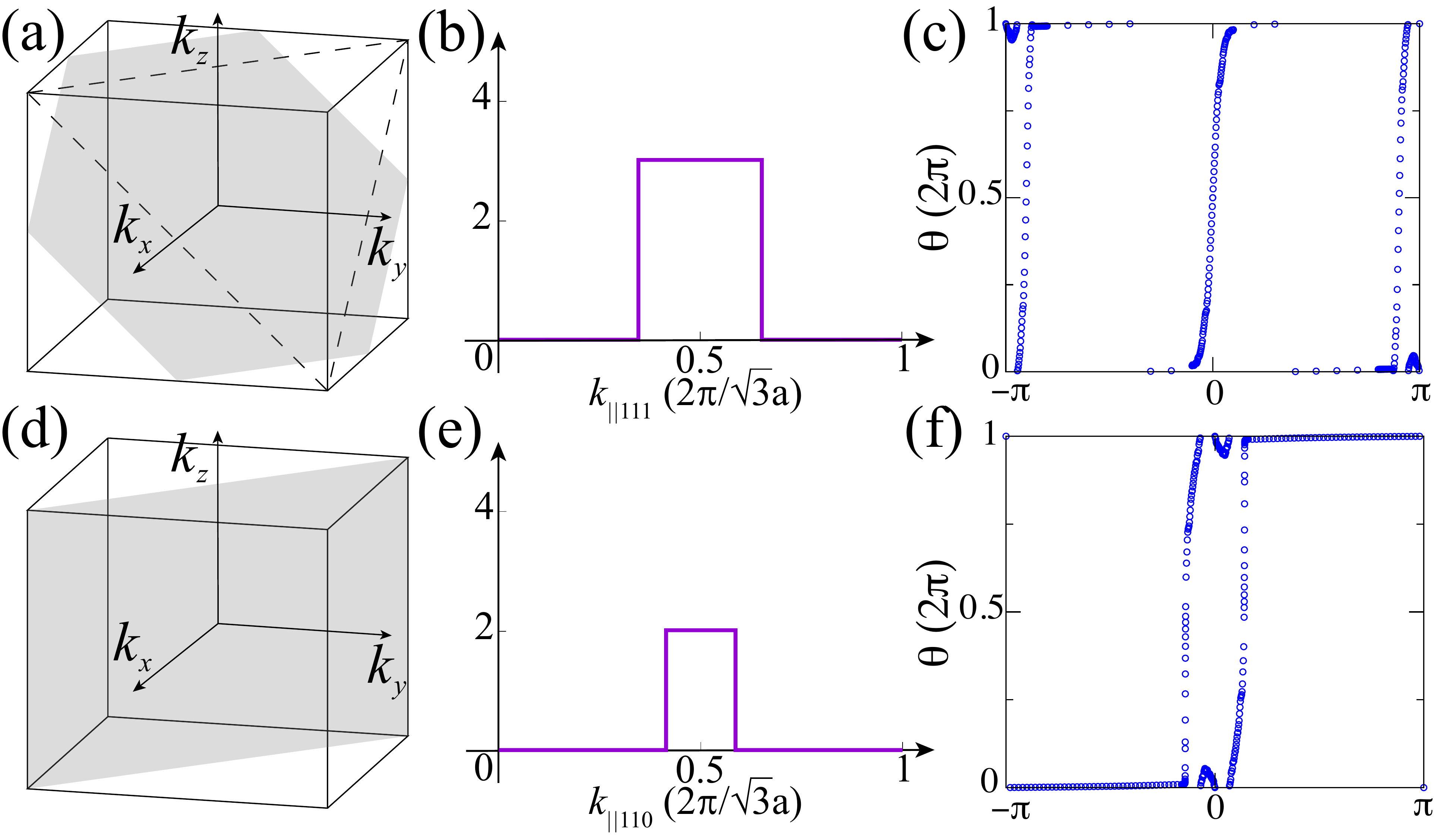}
\caption{ \textbf{(a)} and \textbf{(d)} Two representative (111) and (110) planes in momentum space, respectively. The gray planes and dotted lines indicate 0 and 0.5 displacements, respectively.
 (b) and (e) The evolution of the Chern number as  a function of the displacement of the (111) and (110) planes
 from the $\Gamma$ point, respectively. (c) and (f) The Wilson-loop spectrum for the (111) plane with $k_{||[111]}=0.5~(\frac{2\pi}{\sqrt{3}a})$
 and the (110) plane with $k_{||110}=0.5~(\frac{2\pi}{\sqrt{2}a})$, respectively.
}
\label{ab-chern}
\end{figure}

\subsection{E. Distribution of Weyl nodes}
In the FM2 and FM3 phases, due to the emergence of Weyl nodes, we calculate the Chern numbers of the (111) planes and (110) planes in momentum space as a function of the displacement from the $\Gamma$ point, as shown in Fig. \ref{ab-chern}(a) and Fig. \ref{ab-chern}(d), respectively. The results of the (111) planes are shown in Fig.~\ref{ab-chern}(b). For the gray (111) plane ({\it i.e.} $k_{||[111]}=0$), the Chern number is computed to be 0. By increasing the displacement, the Chern number successively takes two jumps of 3 and -3 when the plane passes three positive and negative Weyl points, respectively. For the (111) plane with $k_{||[111]}=0.5~[\frac{2\pi}{\sqrt{3}a}]$ (denoted by a dashed triangle), its Wilson-loop spectrum is presented in Fig.~\ref{ab-chern}(c), which indicates the corresponding Chern number is 3. After we get the critical displacement, it is easy to find the exact positions of the Weyl nodes. For the (110) plane, similar method is used to find the Weyl nodes, as shown in Fig.~\ref{ab-chern}(e) and (f). At last, the positions of Weyl nodes found in both FM2 and FM3 phases of EuB$_6$ are shown in Table \ref{pos}.

\begin{table}[h]
\caption{The positions of Weyl nodes found in both FM2 and FM3 phases of EuB$_6$.}
\begin{tabular}{c |c c c|c c c |c }
\hline

        Phases      & $~~k_x$ ($\frac{2\pi}{a}$) &~~ $k_y$ ($\frac{2\pi}{a}$)&~~ $k_z$ ($\frac{2\pi}{a}$)      &~~~$k_x$~(\AA$^{-1}$)    &~~~$k_y$~(\AA$^{-1}$)  &~~~$k_z$~(\AA$^{-1}$)& E-E$_F$ (meV)  \\
\hline
                       &  0.0392 & 0.0392 &  0.5+0.0757  &0.0580  & 0.0580 & 0.8517  &  \\
                       \cline{2-7}
                        &-0.0392 &-0.0392 &  0.5-0.0757 &-0.0580  & -0.0580 & 0.6277 &  \\
\cline{2-7}
                       & 0.5+0.0757 & 0.0392 &  0.0392  & 0.8517  & 0.0580 &  0.0580  &  \\
\cline{2-7}
    FM2 phase & 0.5-0.0757 &-0.0392 & -0.0392  &0.6277  & -0.0580 & -0.0580 &  \\
\cline{2-7}
                        & 0.0392  & 0.5+0.0757 &  0.0392  &0.0580  & 0.8517 & 0.0580  & 3.245 \\
\cline{2-7}
                        &-0.0392 & 0.5-0.0757 &-0.0392   &-0.0580  &  0.6277  & -0.0580 &  \\
\hline
                        & 0.5+0.0616 & 0.0608 &   0         &0.8308  &   0.0900 & 0  & \\
\cline{2-7}
                        & 0.0608 & 0.5+0.0616 &   0         &0.0900 &  0.8308 & 0   &  \\
                     \cline{2-7}
FM3 phase      & 0.5-0.0616 & -0.0608 &   0         &0.6485 &  -0.0900  & 0   & -10.272   \\
                     \cline{2-7}
                      & -0.0608  &  0.5-0.0616 &   0         &-0.0900 &   0.6485  & 0 &  \\
\hline
\end{tabular}
\label{pos}
\end{table}

In the FM2 phase, for some specific parameters (such as $u=0.207$ and U = 8 eV), each Weyl node [Fig.~3(c)] can change to three Weyl nodes with invariant total charge, as shown in Fig.~\ref{wzjss}(a).
For example, the negative Weyl point at (0.0392, 0.0392, 0.5747) evolves into a positive and two negative Weyl nodes with the positions shown in Table \ref{weyl3}.
But they are very close to each other, which has little effect on the calculation of AHC (Fig. 4(a)).
As shown in Figs. \ref{wzjss}(b) and (c), the Chern number of the (111) plane successively takes four jumps of -3, 6, -6 and 3 when the plane passes three
negative, six positive, six negative and three positive Weyl points, respectively.

\begin{figure}[h]
\includegraphics[clip,scale=0.38, angle=0]{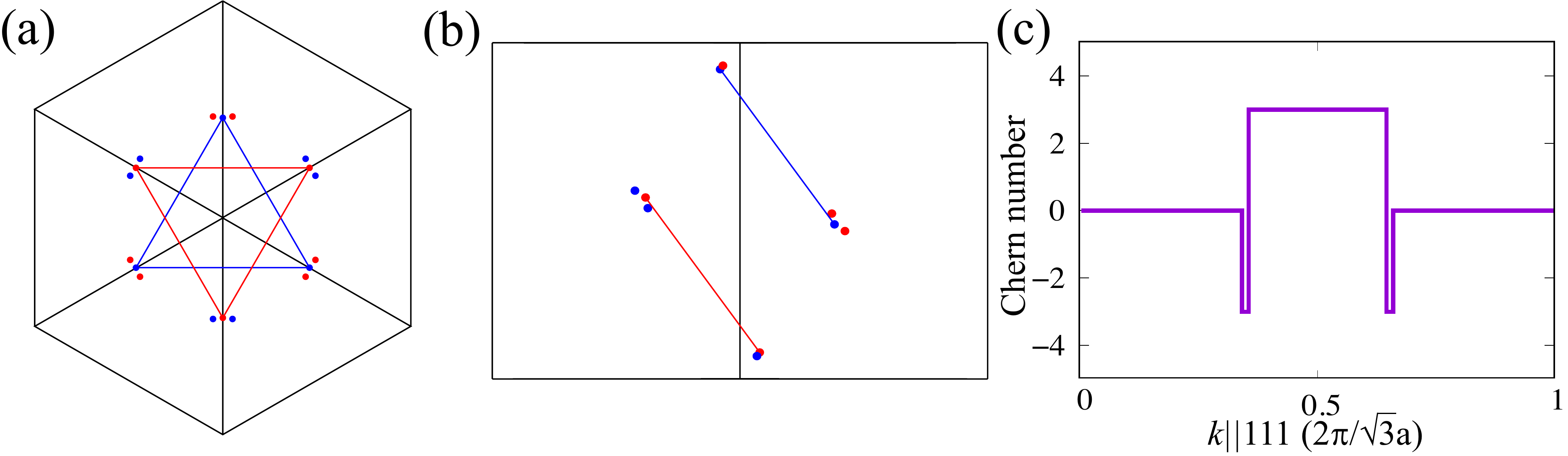}
\caption{ (a) and (b) are the 111- and $\bar{1}$10-view of FM2 EuB$_6$ with $u=0.207$. (c) The evolution of the Chern number as a function of the displacement of the (111) plane from the $\Gamma$ point.}
\label{wzjss}
\end{figure}

\begin{table}[h]
\begin{center}
\caption{The positions of three Weyl nodes coming from the Weyl node at (0.0392, 0.0392, 0.5747) in units of [$2\pi/a$, $2\pi/a$, $2\pi/a$].}
\label{weyl3}
\begin{tabular}{p{2.5cm}|p{2.5cm}|p{2.5cm}|p{2.5cm}}
\hline
\hline
 Weyl node& $k_x$ ($\frac{2\pi}{a}$) & $k_y$ ($\frac{2\pi}{a}$) & $k_z$ ($\frac{2\pi}{a}$) \\
\hline
Weyl node 1  & 0.0399 & 0.0399 & 0.5787 \\
\hline
Weyl node 2  & 0.0083 & 0.0604 & 0.5677 \\
\hline
Weyl node 3  & 0.0604& 0.0083  & 0.5677\\
\hline
\hline
\end{tabular}
\end{center}
\end{table}

\subsection{F. Chern numbers of FM3 EuB$_6$ quantum wells}

Based on the effective $k\cdot p$ models in Section D of the SM, the effective low-energy models for the quantum wells grown along the [111] direction can be
easily obtained by the following two steps. Here we choose $H^{FM2}_Z(k_x, k_y, k_z)$ as an example in the following demonstration. First, we redefine
the coordinate system with three new base vectors $\textbf{1}\equiv (\hat x +\hat y-2 \hat z)/\sqrt 6$, $\textbf{2}\equiv (-\hat x +\hat y)/\sqrt 2$, and $\textbf{3}\equiv (\hat x +\hat y+ \hat z)/\sqrt 3$. Thus, the rotation operator $\hat{R}$ is used to rotate the coordinate system, where $\hat{R}=\begin{pmatrix}
1/\sqrt{6} &  1/\sqrt{6} &-2/\sqrt{6}\\
-1/\sqrt{2} &  1/\sqrt{2} & 0 \\
1/\sqrt{3} &  1/\sqrt{3} &1/\sqrt{3}
\end{pmatrix}$. After the rotation, the \textbf{3}-axis is defined along
the [111] direction. For the momentum vector, we have the equalities $(k_1, k_2, k_3) \cdot \hat{R}=(k_x, k_y, k_z) $, where $k_{1,2,3}$ are defined with respect to the new coordinate system. By replacing $k_{x,y,z}$ with $k_{1,2,3}$, it is easy to get the effective $k\cdot p$ model for the Z point in the
new coordinate system, \emph{i.e.} $H^{FM2}_Z(k_1, k_2, k_3)$.
Next, the open boundary is applied along the $z'$ direction ({\it i.e.} \textbf{3}-axis). The quantum well EuB$_6$ is located in the range $-d/2 \leq z'  \leq d/2$, and vacuum is added in the range $z'<-d/2 $ and $z'>d/2 $, where $d$ is the thickness of the quantum well. So $k_3$ is not a good quantum number and replaced by $-i\partial_{z'}$. Then the Hamiltonian for the quantum well can be written as $H^{FM2}_Z(k_1, k_2, -i\partial_{z'})$.

Next the Chern number is calculated by the Wilson loop method.
The Wilson loops are defined as the circles with the center at the Z point, whose Wannier charge centers (WCCs) are calculated.
By studying the evolution of the WCCs as a function of the radius of the circle ($r$), it is easy to get the
Chern number. For the quantum well with thickness $d=30$ $\text{\AA}$ (60 $\text{\AA}$), the
charge center smoothly shifts upward from 0 to $2\pi$ ($4\pi$), indicating the Chern number $C=1$ (2), as shown in Fig. \ref{chern}.
For FM2 EuB$_6$, X, Y and Z points are equal and can be related by the three-fold rotational symmetry. So they have the same results.
At last, the total Chern numbers for the quantum well with $d=30$ $\text{\AA}$ (60 $\text{\AA}$) is 3 (6), as shown in the Fig. 4(d).

\begin{figure}[h]
\includegraphics[clip,scale=0.25, angle=0]{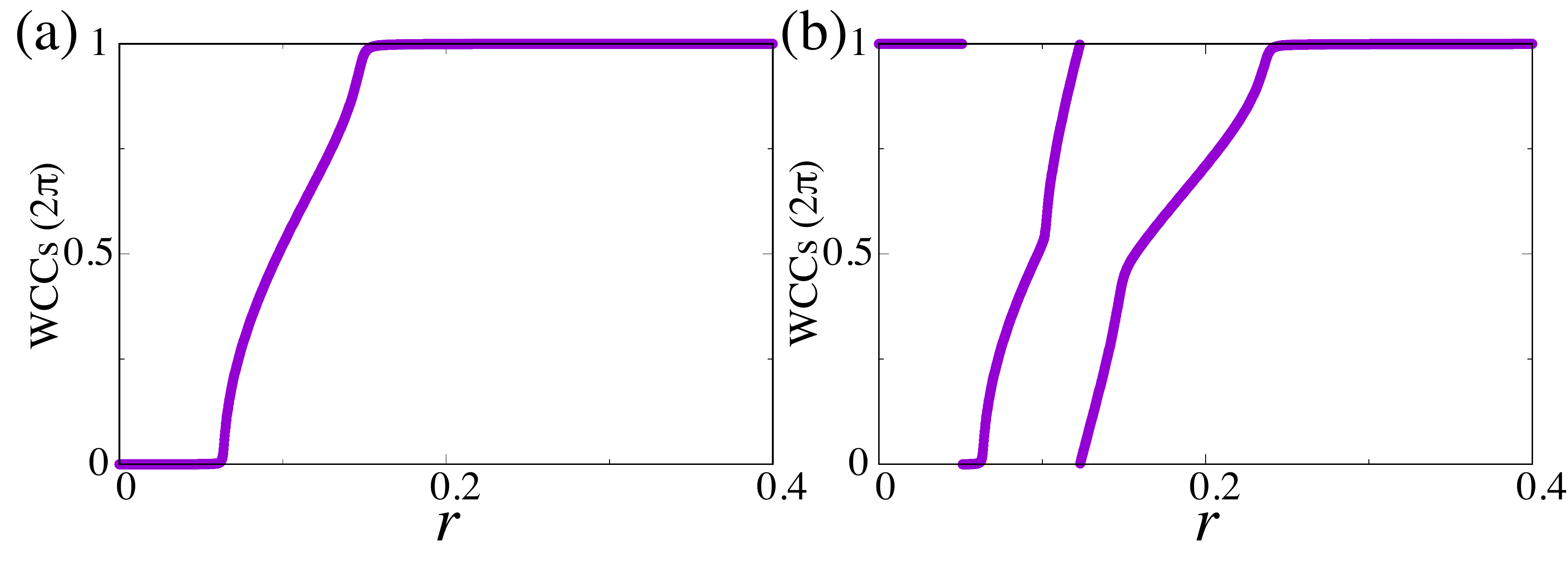}
\caption{(a) and (b) The evolution of WCCs as a function of the radius (r) for the quantum wells with thickness $d=30$ $\text{\AA}$ and 60 $\text{\AA}$, respectively.
}
\label{chern}
\end{figure}

\subsection{G. (001) surface states of FM1 EuB$_6$}
The calculated (001) surface states of FM1 EuB$_6$ is shown in Fig. \ref{001surfstate}, which clearly shows the existence of the drumhead surface states within the energy gap.

\begin{figure}[b]
\includegraphics[clip,scale=0.18, angle=0]{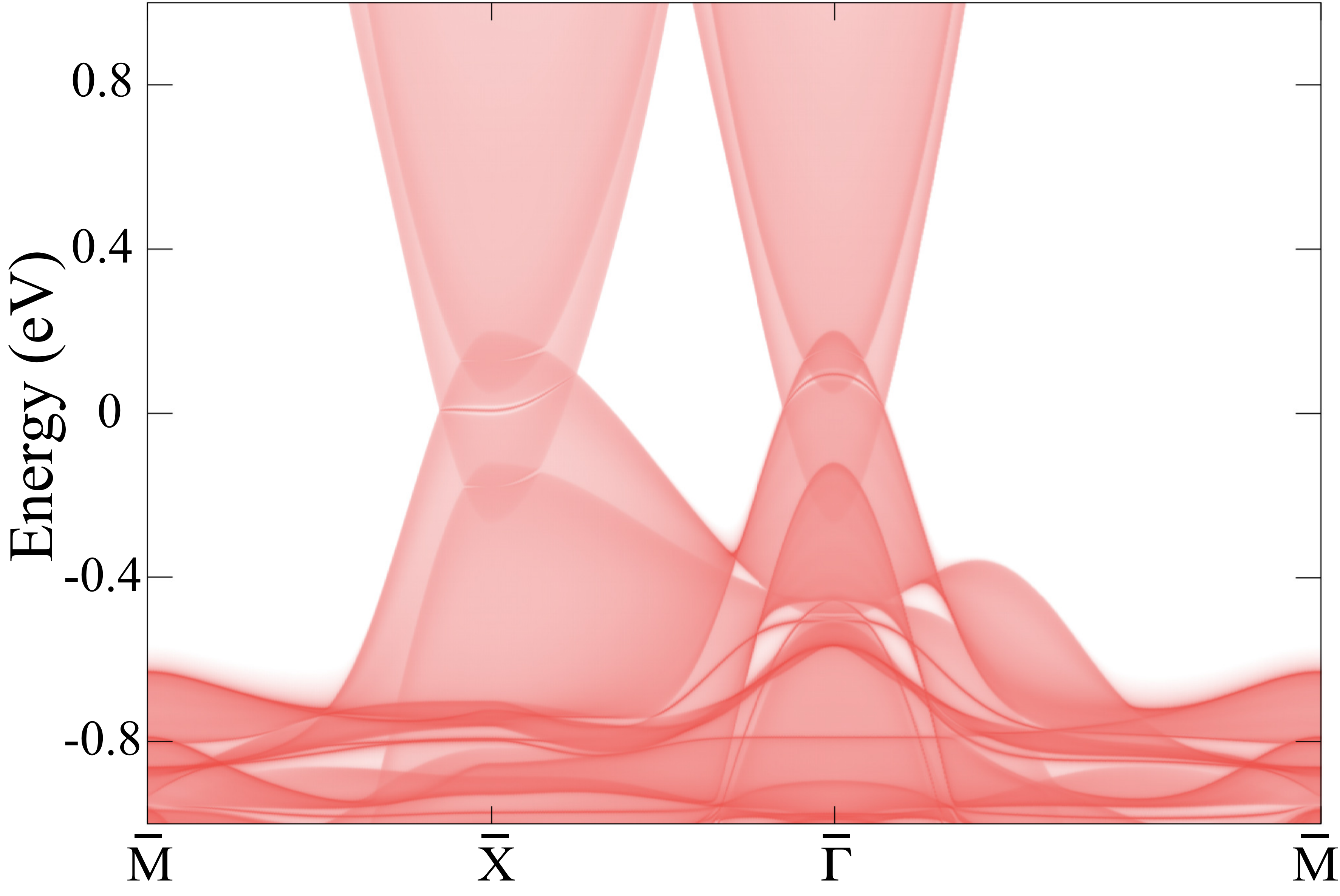}
\caption{The (001) surface states of FM1 EuB$_6$. }
\label{001surfstate}
\end{figure}

\clearpage

\end{widetext}

\end{document}